# Layer shift factor in layered hybrid perovskites – univocal quantitative descriptor of composition-structure-property relationships

Ekaterina I. Marchenko [1,3], Vadim V. Korolev [2,4], Artem Mitrofanov [2,4], Sergey A. Fateev [1], Eugene A. Goodilin [1,2], Alexey B. Tarasov [*1,2]

[1] Laboratory of New Materials for Solar Energetics, Faculty of Materials Science, Lomonosov Moscow State University; 1 Lenin Hills, 119991, Moscow, Russia

[2] Department of Chemistry, Lomonosov Moscow State University; 1 Lenin Hills, 119991, Moscow, Russia

[3] Department of Geology, Lomonosov Moscow State University; 1 Lenin Hills, 119991, Moscow, Russia

[4] Science Data Software, LLC, 14909 Forest Landing Cir, Rockville, MD 20850, United States

**ABSTRACT:** Ascending interest of the scientific community in layered hybrid halide perovskites (LHHPs) as materials for innovative photovoltaic and optoelectronic applications led to unprecedented expansion of this family of compounds, reaching now several hundred refined structures. Despite the unique structural diversity of LHHPs, traditional approaches of describing their structures, such as dividing into Dion-Jacobson (DJ) or Ruddlesden–Popper (RP) phases for most structures are ambiguous and unquantifiable. Here, we introduced a quantitative layer shift factor (LSF) for a univocal classification and quantitative comparison of the structures. We also developed an algorithm for automatic calculation of the LSF for such structures. We demonstrate the application of the proposed approach for an analysis of correlations between the LSF and band gap to reveal "structure-property" relationships. Our study gives a simple and useful approach to classify of either the layered perovskite-like structures or other layered compounds composed of layers of vertex-connected octahedra as a structural unit.

## INTRODUCTION

The family of layered organic-inorganic halide perovskite-like compounds, often referred as "2D hybrid halide perovskites", is derived from the perovskite structural type, exhibits unprecedented structural flexibility which opens prospects for the design of various innovative materials for photovoltaics and optoelectronics. This class of materials shows a set of unique functional properties, such as record-breaking quantum yield of photo- and electroluminescence, tunable narrow emission or broad white-light emission and excellent photoconductivity [1–3].

Layered (2D) hybrid halide perovskites (LHHPs) are resulted of dimentionality reduction of the 3D parent compound along a specific crystallographic plane [4] as reflected by the general formula of $(A`)_{2/q}A_{n-1}B_nX_{3n+1}$ for the most common structural type with dimentionality reduction of (100), where $[A`]^{q+}$ represents singly (q=1) or doubly (q=2) charged organic interlayer (spacer) cation, $A^+$ is a small intralayer cation (such as $Cs^+$, $CH_3NH_3^+$, $[HC(NH_2)_2]^+$); $B^{2+}$ = $Pb^{2+}$, $Sn^{2+}$, $Ge^{2+}$, etc.; $X^-$ = $Cl^-$, $Br^-$, $I^-$; n is the number of layers of corner-shared octahedra within a perovskite slab.

In a recent paper, Saparov and Mitzi [5] categorized the structural types of LHHPs based on their different octahedral connectivity and layer orientation. Particularly, they are subdivided into three main categories: the (100)-oriented, the (110)-oriented, and the (111)-oriented structures based on dimentionality reduction along the specific crystallographic directions. The (100) type of LHHPs was further subdivided, similarly to well-known layered oxide perovskites, into the so-called Dion-Jacobson (DJ) phases and Ruddlesden-Popper (RP) phases. The first is characterized by a relative shift of the neighboring layers along the *ab*-plane (1/2, 0 shift, $KLaNb_2O_7$-type) [6] or no shift at all (0, 0 shift, $CsLaNb_2O_7$-type) [7], while the second type exhibits the (1/2, 1/2) shift [8], as shown on Fig. 1.

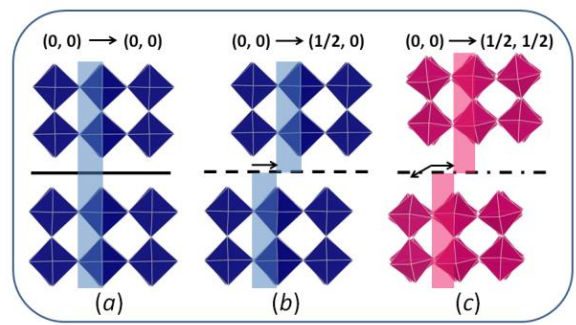

Figure 1. Schematic representation of end members of LHHPs with different stacking of layers: (a) DJ phase with a mirror plane between layers, (b) DJ phase with a plane containing a translation vector (0, ½) along the plane with length equal to a half of the octahedron, (c) RP phase with a plane containing a translation vector (½, ½) along the plane with the same length.

However, the already well-known "DJ-RP classification" has a qualitative, unquantifiable origin and the majority of the structures of known (100) LHHPs can not be strictly attributed to DJ or RP phases. An attempt to eliminate the ambiguity in the definition of the DJ phases was done by Kanatzidis et al. [8] who introduced a new family of layered perovskite structures with alternating cations in the interlayer space (so-called ACI phases). The (110)-oriented structures are usually noted as "eclipsed" and "staggered", while there is no classification of the (111)-oriented structures at all.

It is worth noticing that both experimental data and theoretical calculations indicate that the relative shift of the layers is one of the key factors affecting such important properties of LHHP as band gap, band gap dispersion (charge mobility), and exciton energy, which is excellently illustrated by the significant difference in the optoelectronic properties of RP, DJ and ACI phases [8–10]. Thus, the numerical parameter, characterizing the degree of relative shift of the layers in LHHP is highly required to reveal quantitatively "composition – structure – property" correlations among the whole multitude of the layered perovskites.

Here, we introduce a new quantitative **Layer Shift Factor (LSF)** for univocal classification and comparison of the layered perovskite structures. We also present a simple Python script for automatic calculation of the LSF for any structure. This parameter was used for crystal-chemical analysis to establish relationships between the composition, structural features and their relationship with functional properties.

## METHODS

Crystal structures of the known layered (100)- and (110)-hybrid perovskites for the analysis were taken from the recently published database of 2D perovskite-like materials [11]. Only the structures with integer occupancies of crystallographic positions were used for the calculations. The doubled direction in one of three supercells, 2×1×1, 1×2×1, and 1×1×2, was defined as a vector of two-dimensional periodicity where the calculated number of connected components is exactly twice as high as the number of connected vertex-connected octahedral (Figure S6 in SI). Atomic connectivity is determined then based on the interatomic distances, namely, two atoms are considered as bonded if the distance between them does not exceed a sum of corresponding Cordero covalent radii [12]. A user-defined tolerance distance has a typical value of 0.25 Å [13,14] while for some structures with a distorted inorganic framework it was manually set equal to 0.85 Å.

The **Layer Shift Factor (LSF)** was calculated for the examinated structures as a vector $D$ $(t_1, t_2)$ - where $t_1$ and $t_2$ are translational vectors of shift reaching maximum values equal to a half of the octahedron. LSF was calculated by the following algorithm. Metal/metalloid atoms included into two adjacent perovskite octahedral layers were examined. For each possible pair of atoms $(a_i, a_j)$, where $a_i$ and $a_j$ belong to the first and second layers, respectively, vector $\bar{a}_i - \bar{a}_j$ was calculated. This vector was projected onto the plane formed by the two crystallographic axes lying in the perovskite plane (Figure S7 in SI). The projection of the shortest vector required to bring the octahedral layers to zero in-plane displacement was defined as the first coordinate of the translation vector $D$ (translation 1); the second coordinate of D (translation 2), for the sake of simplicity, was defined as a vector perpendicular to the first one and lying in the plane of two-dimensional periodicity. Strictly speaking, in some structures, a slight asymmetry of octahedra is observed; in this case, the translation unit vectors are not perpendicular. However, this simplification does not lead to a significant mismatch in the value of in-plane displacement. Particularly, the DJ and RP phases among (100) perovskites are the end members of possible LSF values - (0,0) / (½,0) and (½, ½), respectively.

A series of hypothetical (100) and (110) structures with the same composition $Cs_2PbBr_4$ and different LSF were generated manually. One-layered perovskite slabs consisting of non-distorted corner-shared $PbBr_6$ octahedra were spaced 3.5 Å and 4.6 Å apart respectively, all the Pb-Br distances were chosen to be 2.98 Å. The Cs atoms were placed at the center of mass of the eight nearest Br atoms. A two-dimensional grid with a step of 0.1 was used to generate 21 structures with different LSF parameter values. These all-inorganic hypothetical structures can serve as relevant prototypical models of 2D hybrid perovskites for DFT band structure calculations since the substitution of interlayer molecular cations by cesium does not alter significantly the band structure near the electronic gap[15,16].

For the generated set of $Cs_2PbBr_4$ structures the band gaps were calculated using density functional theory (DFT) as implemented in the GPAW [17,18] code. The projector-augmented wave (PAW) formalism [19,20] with a cutoff energy of 500 eV was used to describe the electronic wavefunctions. We used a Monkhorst–Pack Γ-centered grid [21] with the *k*-point density of 10 points/Å for sampling the Brillouin zone. The band gaps were calculated using the GLLB-SC exchange-correlation functional [22,23], including spin-orbit coupling as described in ref. [24].

## RESULTS AND DISCUSSION

In total, LSF was calculated for 282 (100) and 21 (110) layered hybrid perovskite structures (Figure 2 and Figure 3) and it can be seen that in contrast to dual RP-DJ and "eclipsed - staggered" classifications, all analyzed structures encompass whole configuration space of vector LSF $(t_1, t_2)$. Despite the common believe that the structures with diammonium cations tend to form predominantly DJ phases (LSF (0, 0)) and that monoammonium cations are mainly associated with the structures with RP phases (LSF (1/2, 1/2))[25], in terms of LSF we clearly see no systematic correlation between cation types and layer shift (see Figures S1-S3 in SI). It should be noticed that while LSF is a parameter derived from the crystal structure, it is sensitive for phase transitions and therefore is a function of temperature (see Fig. S5 in SI)

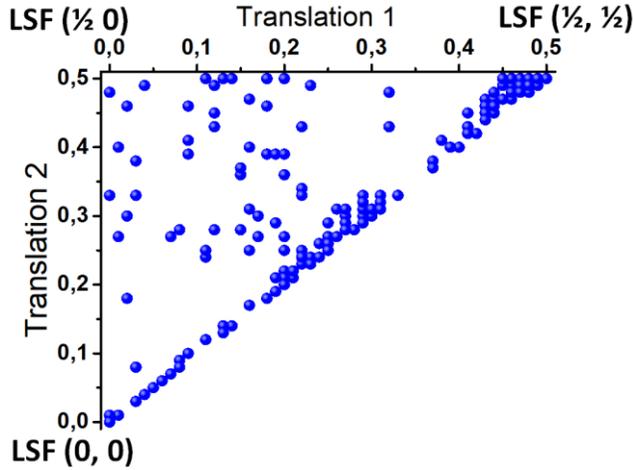

Figure 2. Calculated layer shift factor ($t_1, t_2$) for (100) LHHPs from the Database [11]. DJ and RP phases correspond to the vectors (0, 0), (1/2, 0) and (1/2, 1/2), respectively. Translations 1 and 2 are $t_1$ and $t_2$, respectively.

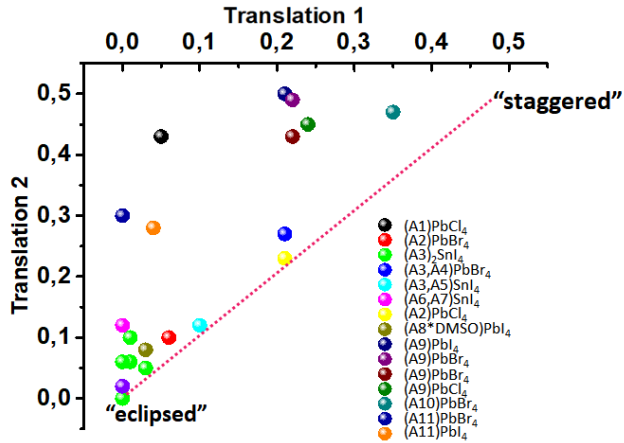

Figure 3. Calculated layer shift factor ($t_1, t_2$) for (110) layered hybrid perovskite-like structures from the Database[11]. "Eclipsed" and "staggered" conformations correspond to the vectors (0, 0) and (1/2, 1/2), respectively. Organic cations are designated as follows: A1 – piperazinium, A2 – N-propylammonio-3-imidazolium, A3 – guanidinium, A4 – 1,2,4-triazolium, A5 – 1-methylimidazolium, A6 – iodoformamidinium, A7 – methylammonium, A8 – ethanediammonium, A9 – 3-ammonioimidazolium, A10 – N-methylethane-1,2-diammonium, A11 – 2,2'-(ethylenedioxy)bis-(ethylammonium).

To reveal possible influence of cationic composition on structural features of LHHPs, we analyzed the homologous series of single-layered (n=1) iodoplumbate perovskites $A_2PbI_4$ with interlayer alkylammonium cations $A^+$ of different carbon chain length (from ethylammonium to decylammonium)[11]. These $A_2PbI_4$ perovskites can adopt orthorhombic space group symmetry $Pbca$ or monoclinic one $P2_1/c$. The structures with the orthorhombic symmetry belong to the LSF (½, ½) (RP) while the monoclinic structures are intermediate between LSF (0, 0) (DJ) and LSF (½, ½) (RP) phases within a nearly linear trend (Figure 4a). Such a difference in LSF values may be caused by the existence of different conformations of interlayer cations corresponding to a different penetration depth of these cations into the inorganic layer of corner-shared octahedra as shown in Figure 4b. For monoclinic structures of this series, there is a large distribution in penetration depth values, than that, in contrast, observed for orthorhombic structures with LSF close to (1/2, 1/2).

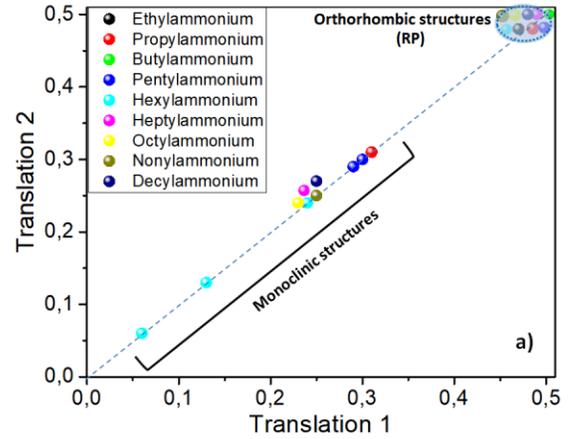

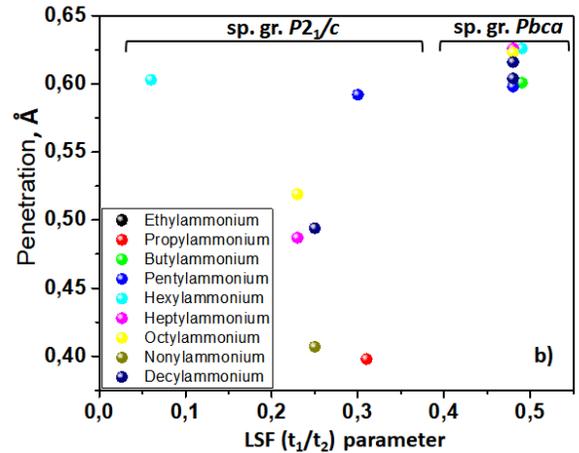

Figure 4. Calculated values of LSF (a) and penetration depth of interlayer cations ($A^+$) into the inorganic layer (b) for (100) single-layered (n=1) $A_2PbI_4$ LHHPs with alkylammonium cations of different length.

The penetration depth of organic cations into inorganic layers correlates with Pb-X-Pb angles distortion and, in turn, with band gaps for single-layered halide perovskites [11]. In order to reveal an influence of LSF values on the band gaps, they were calculated for the series of hypo-thetical structures of (100) and (110) single-layered bromoplumbate perovskites ($Cs_2PbBr_4$) characterized by undistorted inorganic framework with the *P*4/*mmm* symmetry and a small interlayer distance of 3.5 Å and 4.6 Å respectively. For hypothetical series of single-layered bromoplumbate perovskites the calculated band gap values monotonically increase with an increase of the LSF values from (0, 0) to (½, ½) and from (0, 0) to (½, 0) for (100) and (110) types of structures respectively, as shown in Figure 5 and Figure S4. This dependency is explained by the decreasing of the overlapping of axial p-orbitals for terminal halogen atoms of adjacent inorganic layers. This opens the way for rational search of organic cations which would form the structures with the highest (LSF (½, ½)) or the lowest (LSF (0, 0)) bang gap values.

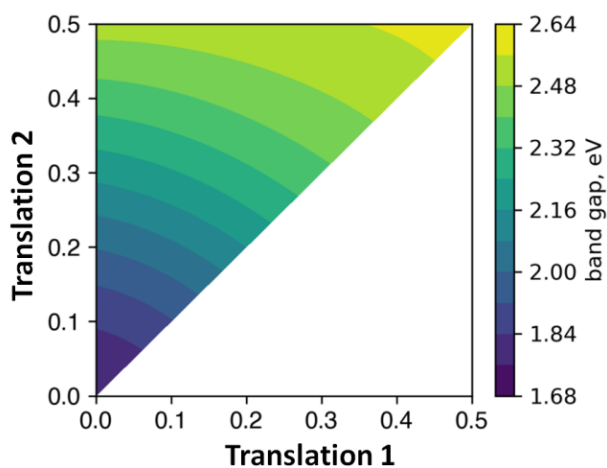

Figure 5. Calculated band gaps of the series of hypothetical $Cs_2PbBr_4$ single-layered (100) perovskites with the interlayer distance of 3.5 Å as a function of the LSF ($T_1$, $T_2$).

## CONCLUSIONS

We proposed a new, universal and quantitative, layer shift factor (LSF) for a quantitative comparison and univocal classification of LHHPs and developed a simple Python-based script for its calculation. We revealed that the orthorhombic symmetry structures with alkylammonium cations belong to the "pure" RP phase while the monoclinic structures are linearly distributed between DJ (LSF = (0, 0)) and RP (LSF = (1/2, 1/2)) phases. Such different LSF values are related to the existence of different conformations of interlayer cations. We show that the LSF parameter correlates well with the penetration depth of organic cations into inorganic layers, affecting therefore the distortion of inorganic framework and the band gap values. Therefore, the suggested simple layer shift factor may be considered as a key parameter for rational crystal chemical analysis and revealing composition-structure-property correlations for these materials. The discussed approach strengthens our understanding of the origin of "layered perovskites" and highlights fundamental questions which can be considered using the methodology presented in this work.

## ASSOCIATED CONTENT

The algorithm of program is available on https://github.com/korolewa/lsc

The Supporting Information contains Layer shift factor values for (100) layered perovskite structures, calculated band gaps for (110) layered perovskite structures, schemes of LSF calculation. This material is available free of charge via the Internet at http://pubs.acs.org.

## AUTHOR INFORMATION

**Corresponding Author**

* alexey.bor.tarasov@yandex.ru

**Author Contributions**

The manuscript was written through contributions of all authors. All authors have given approval to the final version of the manuscript.

**Notes**

The authors declare no competing financial interest.

## ACKNOWLEDGMENT

This work was financial supported by a grant from the Russian Science Foundation, project number 19-73-30022.

## REFERENCES

(1) Mao, L.; Stoumpos, C. C.; Kanatzidis, M. G. Two-Dimensional Hybrid Halide Perovskites: Principles and Promises. *J. Am. Chem. Soc.* **2019**, *141* (3), 1171–1190. https://doi.org/10.1021/jacs.8b10851.

(2) Katan, C.; Mercier, N.; Even, J. Quantum and Dielectric Confinement Effects in Lower-Dimensional Hybrid Perovskite Semiconductors. *Chem. Rev.* **2019**, *119* (5), 3140–3192. https://doi.org/10.1021/acs.chemrev.8b00417.

(3) Vashishtha, P.; Ng, M.; Shivarudraiah, S. B.; Halpert, J. E. High Efficiency Blue and Green Light-Emitting Diodes Using Ruddlesden-Popper Inorganic Mixed Halide Perovskites with Butylammonium Interlayers. *Chem. Mater.* **2019**, *31* (1), 83–89. https://doi.org/10.1021/acs.chemmater.8b02999.

(4) Wang, J.; Dong, J.; Lu, F.; Sun, C.; Zhang, Q.; Wang, N. Two-Dimensional Lead-Free Halide Perovskite Materials and Devices. *J. Mater. Chem. A* **2019**, *7* (41), 23563–23576. https://doi.org/10.1039/c9ta06455a.

(5) Saparov, B.; Mitzi, D. B. Organic–Inorganic Perovskites: Structural Versatility for Functional Materials Design. *Chem. Rev.* **2016**, *116* (7), 4558–4596. https://doi.org/10.1021/acs.chemrev.5b00715.

(6) Sato, M.; Abo, J.; Jin, T.; Ohta, M. Structure Determination of KLaNb2O7 Exhibiting Ion Exchange Ability by X-Ray Powder Diffraction. *Solid State Ionics* **1992**, *51* (1–2), 85–89. https://doi.org/10.1016/0167-2738(92)90348-S.

(7) Kumada, N.; Kinomura, N.; Sleight, A. W. CsLaNb2O7. *Acta Crystallogr. Sect. C Cryst. Struct. Commun.* **1996**, *52* (Pt 5), 1063–1065. https://doi.org/10.1107/S0108270195015848.

(8) Soe, C. M. M.; Stoumpos, C. C.; Kepenekian, M.; Traoré, B.; Tsai, H.; Nie, W.; Wang, B.; Katan, C.; Seshadri, R.; Mohite, A. D.; et al. New Type of 2D Perovskites with Alternating Cations in the Interlayer Space, (C(NH2)3)(CH3NH3)NPbnI3n+1: Structure, Properties, and Photovoltaic Performance. *J. Am. Chem. Soc.* **2017**, *139* (45), 16297–16309. https://doi.org/10.1021/jacs.7b09096.

(9) Nazarenko, O.; Kotyrba, M. R.; Wörle, M.; Cuervo-Reyes, E.; Yakunin, S.; Kovalenko, M. V. Luminescent and Photoconductive Layered Lead Halide Perovskite Compounds Comprising Mixtures of Cesium and Guanidinium Cations. *Inorg. Chem.* **2017**, *56* (19), 11552–11564. https://doi.org/10.1021/acs.inorgchem.7b01204.

(10) Mao, L.; Ke, W.; Pedesseau, L.; Wu, Y.; Katan, C.; Even, J.; Wasielewski, M. R.; Stoumpos, C. C.; Kanatzidis, M. G. Hybrid Dion–Jacobson 2D Lead Iodide Perovskites. *J. Am. Chem. Soc.* **2018**, *140* (10), 3775–3783. https://doi.org/10.1021/jacs.8b00542.

(11) Marchenko, E. I.; Fateev, S. A.; Petrov, A. A.; Korolev, V. V.; Mitrofanov, A.; Petrov, A. V.; Goodilin, E. A.; Tarasov, A. B. Database of Two-Dimensional Hybrid Perovskite Materials: Open-Access Collection of Crystal Structures, Band Gaps, and Atomic Partial Charges Predicted by Machine Learning. *Chem. Mater.* **2020**, *32* (17), 7383–7388. https://doi.org/10.1021/acs.chemmater.0c02290.

(12) Cordero, B.; Gómez, V.; Platero-Prats, A. E.; Revés, M.; Echeverría, J.; Cremades, E.; Barragán, F.; Alvarez, S. Covalent Radii Revisited. *J. Chem. Soc. Dalt. Trans.* **2008**, No. 21, 2832–2838. https://doi.org/10.1039/b801115j.

(13) Isayev, O.; Oses, C.; Toher, C.; Gossett, E.; Curtarolo, S.; Tropsha, A. Universal Fragment Descriptors for Predicting Properties of Inorganic Crystals. *Nature Communications*. Nature Publishing Group 2017, pp 1–12. https://doi.org/10.1038/ncomms15679.

(14) Xie, T.; Grossman, J. C. Crystal Graph Convolutional Neural Networks for an Accurate and Interpretable Prediction of Material Properties. *Phys. Rev. Lett.* **2018**, *120* (14), 145301.

(15) Traore, B.; Pedesseau, L.; Assam, L.; Che, X.; Blancon, J.-C.; Tsai, H.; Nie, W.; Stoumpos, C. C.; Kanatzidis, M. G.; Tretiak, S.; et al. Composite Nature of Layered Hybrid Perovskites: Assessment on Quantum and Dielectric Confinements and Band Alignment. *ACS Nano* **2018**, *12* (4), 3321–3332.

(16) Fu, Y.; Jiang, X.; Li, X.; Traore, B.; Spanopoulos, I.; Katan, C.; Even, J.; Kanatzidis, M. G.; Harel, E. Cation Engineering in Two-Dimensional Ruddlesden-Popper Lead Iodide Perovskites with Mixed Large A-Site Cations in the Cages. *J. Am. Chem. Soc.* **2020**, *142* (8), 4008–4021. https://doi.org/10.1021/jacs.9b13587.

(17) Mortensen, J. J.; Hansen, L. B.; Jacobsen, K. W. Real-Space Grid Implementation of the Projector Augmented Wave Method. *Phys. Rev. B* **2005**, *71* (3), 35109.

(18) Enkovaara, J. E.; Rostgaard, C.; Mortensen, J. J.; Chen, J.; Dułak, M.; Ferrighi, L.; Gavnholt, J.; Glinsvad, C.; Haikola, V.; Hansen, H. A.; et al. Electronic Structure Calculations with GPAW: A Real-Space Implementation of the Projector Augmented-Wave Method. *J. Phys. Condens. Matter* **2010**, *22* (25), 253202.

(19) Blöchl, P. E. Projector Augmented-Wave Method. *Phys. Rev. B* **1994**, *50* (24), 17953.

(20) Kresse, G.; Joubert, D. From Ultrasoft Pseudopotentials to the Projector Augmented-Wave Method. *Phys. Rev. B* **1999**, *59* (3), 1758.

(21) Monkhorst, H. J.; Pack, J. D. Special Points for Brillouin-Zone Integrations. *Phys. Rev. B* **1976**, *13* (12), 5188.

(22) Gritsenko, O.; van Leeuwen, R.; van Lenthe, E.; Baerends, E. J. Self-Consistent Approximation to the Kohn-Sham Exchange Potential. *Phys. Rev. A* **1995**, *51* (3), 1944.

(23) Kuisma, M.; Ojanen, J.; Enkovaara, J.; Rantala, T. T. Kohn-Sham Potential with Discontinuity for Band Gap Materials. *Phys. Rev. B* **2010**, *82* (11), 115106.

(24) Olsen, T. Designing In-Plane Heterostructures of Quantum Spin Hall Insulators from First Principles: 1 T′-MoS2 with Adsorbates. *Phys. Rev. B* **2016**, *235106* (23), 1–9. https://doi.org/10.1103/PhysRevB.94.235106.

(25) Tremblay, M.-H.; Bacsa, J.; Zhao, B.; Pulvirenti, F.; Barlow, S.; Marder, S. R. Structures of (4-Y-C 6 H 4 CH 2 NH 3 ) 2 PbI 4 {Y = H, F, Cl, Br, I}: Tuning of Hybrid Organic Inorganic Perovskite Structures from Ruddlesden–Popper to Dion–Jacobson Limits. *Chem. Mater.* **2019**, *31* (16), 6145–6153. https://doi.org/10.1021/acs.chemmater.9b01564.

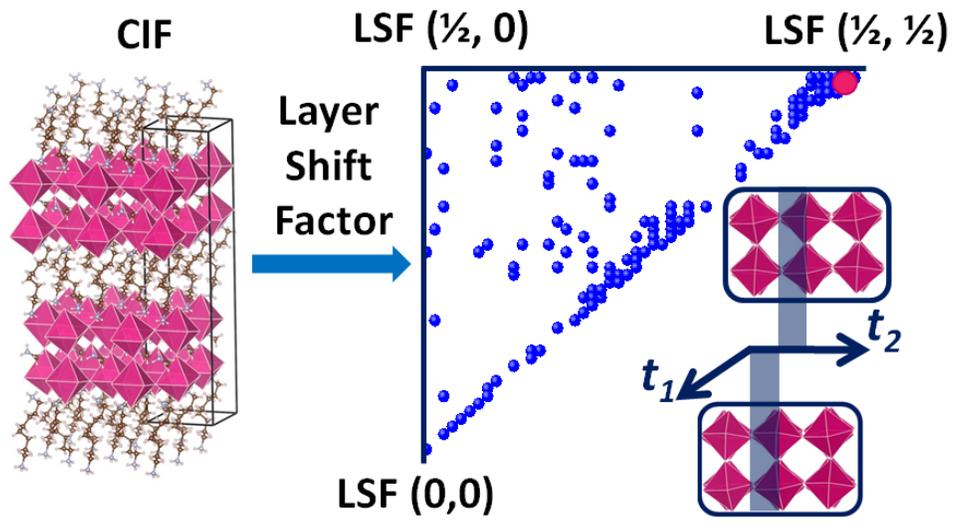